\documentclass[epj]{svjour}
\usepackage{graphics}
\usepackage{multirow}
\usepackage[fleqn]{amsmath}
\usepackage{subfigure}  
\usepackage{float, subfig}
\usepackage{graphicx}
\usepackage{url}

\begin{document}
\title{Photon Reconstruction Performance at the CEPC baseline detector}
\author{Yuqiao Shen\inst{1}, Hong Xiao\inst{1}, Hengmei Li\inst{1}, Sai Qin\inst{1}, Zhen Wang\inst{1}, Changying Wang\inst{1}, Desheng Zhang\inst{1}, Manqi Ruan\inst{2}% etc
\thanks{\emph{e-mail:} Manqi.ruan@ihep.ac.cn}%
}                     % Do not remove
\institute{Changzhou Institute of Technology \and  Institute of High Energy Physics}
\date{Received: date / Revised version: date}
\abstract{
The Circular Electron Positron Collider (CEPC) is a proposed Higgs/Z factory. The photon reconstruction is critical to its physics program. We study the photon reconstruction at the CEPC baseline detector, a Particle Flow oriented detector. We characterize the objective performance on both single-photon and di-photon samples. Using the single-photon samples, we quantify the photon conversion rate, the differential reconstruction efficiency and energy resolution, and the identification performance. Using di-photon samples, our analysis shows that the CEPC baseline detector reaches a relative mass resolution of 1.7 - 2.2\% of the Higgs boson at the $H\to\gamma\gamma$ sample, and can reconstruct the $\pi^0$ with energy as high as 20 - 30 GeV. About 97\% of the $\pi^0$ generated in $Z\to\tau\tau$ events can be reconstructed successfully. We also investigate the impact of geometry defects on photon energy resolution and discuss the possible corrections according to the reconstructed photon position.
\PACS{
      {} {  }
     } % end of PACS codes
} %end of abstract
\maketitle

\section{Introduction}

The discovery of Higgs boson~\cite{a}\cite{b} not only completes the Standard Model (SM) particle spectrum but also provides a very sensitive probe to the fundamental physics principles underlying the SM. A precise Higgs factory becomes a must for the future high energy physics explorations, which may shed light on the profound problems, such as the CP violation, the naturalness, the hierarchy, and the dark matter candidate~\cite{sm1}\cite{sm2}\cite{sm3}\cite{sm4}\cite{sm5}\cite{sm6}. 

The CEPC is a future large scale collider complex. According to its CDR~\cite{cdr}, the CEPC has a main ring circumference of 100 km and two interaction points. It can be operated as a Higgs factor ($\sqrt{s} = 240$ GeV), a Z factory ($\sqrt{s} = 91.2$ GeV) and a W factory ($\sqrt{s} = 160$ GeV). In its Higgs run, the CEPC has a nominally integrated luminosity of 5.6 ab$^{-1}$ and is expected to produce one million Higgs bosons.  Compared to the LHC,  the CEPC has a much cleaner collision environment, well known and adjustable initial state. It provides critical information on the nature of the Higgs boson on top of the HL-LHC measurements. It determines model independently the Higgs total width and couplings. For many critical measurements, such as the Higgs invisible decays and exotic decays~\cite{zl}, the accuracy of CEPC is superior to the HL-LHC by more than one order of magnitude.  After the electron-positron collision phase,  a super proton-proton collider (SPPC) with center-of-mass energy around 100 TeV can be installed in the same tunnel.
 
Photon reconstruction is essential for the CEPC physics measurement and is a critical benchmark for the CEPC detector design and optimization. This paper presents the photon reconstruction performance at the CEPC baseline detector, in terms of the photon conversion rate, the reconstruction efficiency, photon identification, and energy resolution. The $H\to\gamma\gamma$ signal and the $\pi^0$ reconstruction are also analyzed using the corresponding samples.  

This paper is organized as follows. Section~\ref{sec:detectorsoft} introduces the detector geometry, software, and samples used in the studies. Section~\ref{sec:sinpho} describes a performance analysis on single-photon events. The performance on di-photon events is presented in Section~\ref{sec:dipho}. In Section~\ref{sec:con}, a conclusion and a general discussion is summarized.

%+++++++++++++++++++++++Baseline detector, software chain and reconstruction++++++++++++++++++++++++++++++

\section{Baseline detector, software, and samples}
\label{sec:detectorsoft}
\subsection{Baseline detector}
A Particle Flow oriented detector design is selected as the baseline for the CEPC. The detector is composed of a silicon pixel vertex, a silicon inner tracker (SIT), a Time Projection Chamber (TPC), a silicon external tracker (SET), a silicon octagon sampling Electromagnetic Calorimeter (ECAL), a steel-glass Resistive Plate Chamber (RPC) sampling Hadronic Calorimeter (HCAL), a  3 Tesla superconducting  solenoid  and a  flux return yoke embedded with muon detectors.  More information about the geometry can be found in Ref.~\cite{cdr}. The baseline reconstructs all the visible final state particles in their most-suited detector subsystems. Therefore, photon reconstruction provides a strong motivation for ECAL geometry optimization.

\begin{figure}
\begin{minipage}{\textwidth}
\centering 
\includegraphics[width=0.495\textwidth]{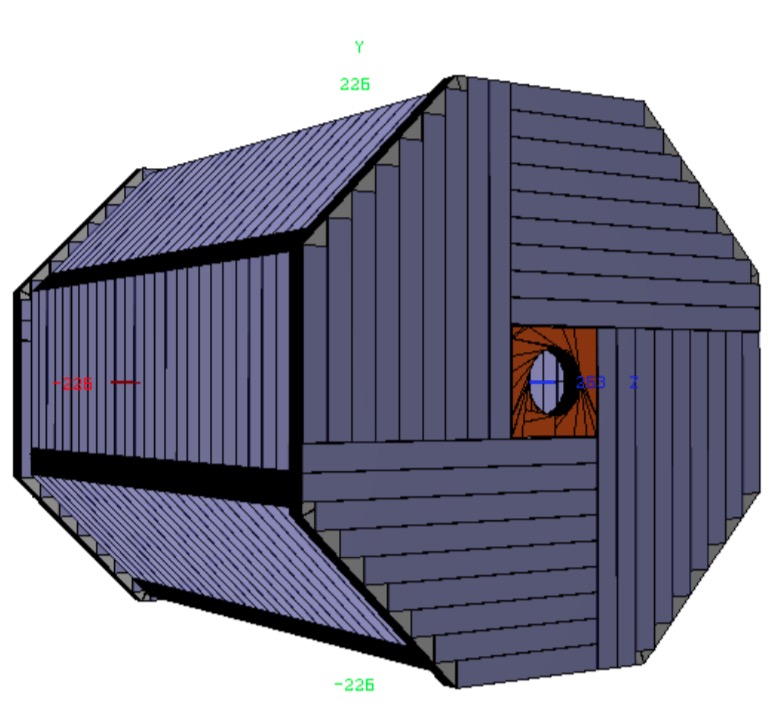}
\end{minipage}
\caption{Schematic of the CEPC ECAL layout in its baseline design. The ECAL is composed of one cylindrical barrel and two disk-like endcap sections. The barrel section is made of 8 staves. Each stave is organized into 5 trapezoidal modules. Each module contains 5 columns. The barrel module is shown in Figure~\ref{fig:ecal2}. The radius of the barrel section is 2028 mm. The two endcap sections are composed of 8 quadrants. Each quadrant is made of 2 modules. Each endcap module contains 7 columns.  The two endcap sections are located at $\pm2635$ mm.}
\label{fig:ecal} 
\end{figure}

\begin{figure}
\begin{minipage}{\textwidth}
\centering 
\includegraphics[width=0.495\textwidth]{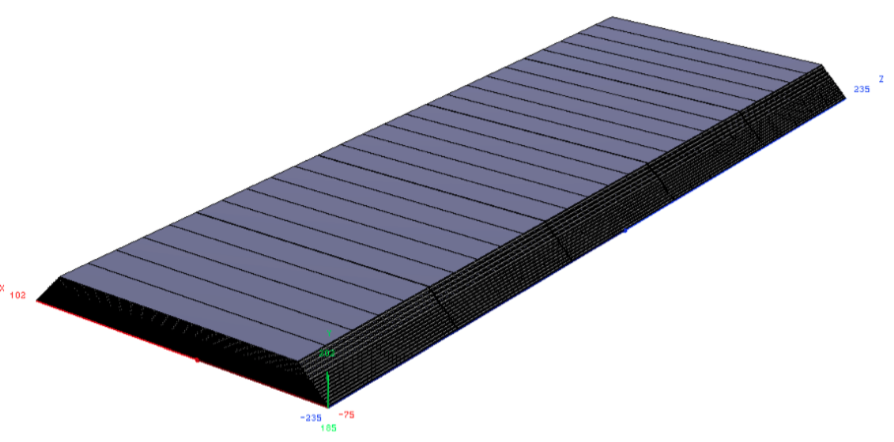}
\end{minipage}
\caption{Schematic of the structure of one ECAL stave.  Each stave is organized into 5 trapezoidal modules. Each module contains 5 columns.}
\label{fig:ecal2} 
\end{figure}

The ECAL is composed of one cylindrical barrel and two disk-like endcap sections as shown in Figure~\ref{fig:ecal}. The barrel section is made of 8 staves. Each stave is organized into 5 trapezoidal modules. Each module contains 5 columns. The barrel module is shown in Figure~\ref{fig:ecal2}. The radius of the barrel section is 2028 mm. The two endcap sections are composed of 8 quadrants. Each quadrant is made of 2 modules. Each endcap module contains 7 columns. The two endcap sections are located at $\pm2635$ mm.
Both the ECAL barrel and endcap are segmented into 30 longitudinal layers with a total thickness of tungsten of 84 mm (corresponding to a total depth of $24X_{0}$). The 30 ECAL longitudinal layers are split into 2 sections with different thickness of absorber layers. The first section contains 20 layers of 2.1 mm thick tungsten plates while the second contains 10 layers of 4.2 mm tungsten plates. The ECAL starts with a sensitive layer. Each sensitive layer is equipped with 0.5 mm thick silicon sensors. The silicon sensor size is $10\times10$ mm$^2$~\cite{cdr}.

%==================software chain======================

\begin{figure}
\begin{minipage}{\textwidth}
\centering
\includegraphics[width=0.495\textwidth]{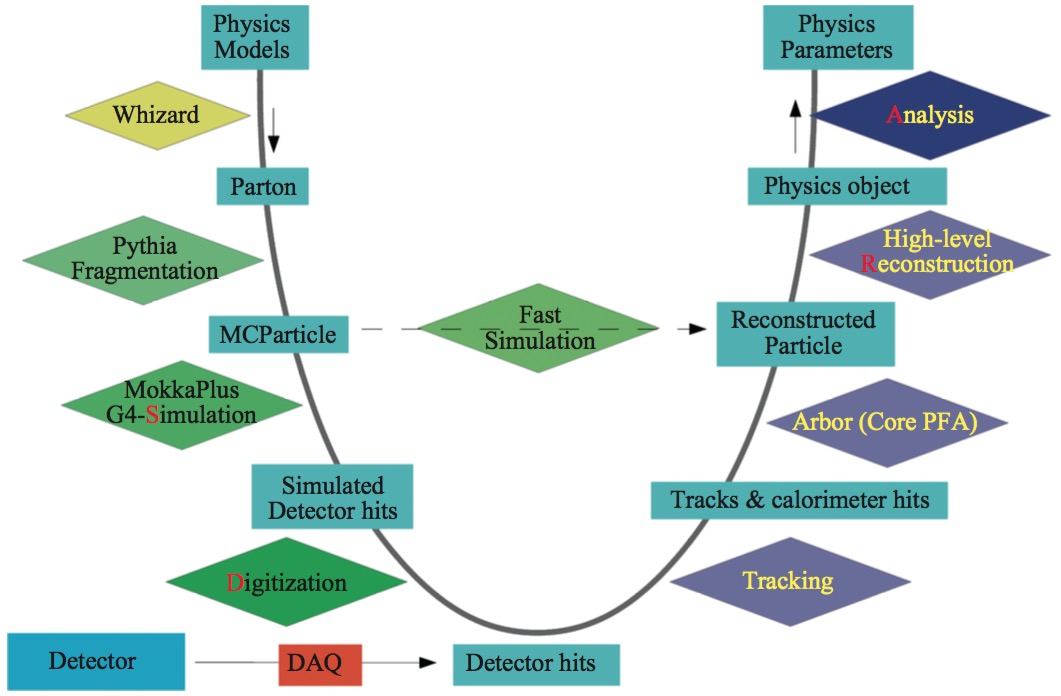}
\end{minipage}
\caption{The flow chart of the CEPC simulation studies. Whizard and Pythia are used to generate final state particles for physics processes. MokkaPlus is used for Simulation. Arbor is used to reconstruct physics objects for further analysis.}
\label{fig:softchain}
\end{figure}

\subsection{Software} 
A complete simulation-reconstruction software has been established for the CEPC baseline detector. Figure~\ref{fig:softchain} shows the major processes of the software chain~\cite{s1}. Whizard~\cite{soft1} and Pythia~\cite{soft3} are used as generator. MokkaPlus~\cite{s2}\cite{s3}, a Geant4~\cite{sadd} based detector simulation framework, is used for the full detector simulation. Arbor~\cite{s5} is developed for the PFA~\cite{s4} reconstruction.

Arbor is inspired by the fact that the particle shower spatial configuration naturally follows a tree configuration. It emphasizes the reconstruction and separation of particle showers induced by the final state particle. It builds calorimeter clusters from calorimeter hits and interprets the clusters and tracks into final state particles. 

\begin{figure}
\begin{minipage}{\textwidth}
\centering 
  \subfigure[0.1 GeV photon]{\includegraphics[width=0.495\textwidth]{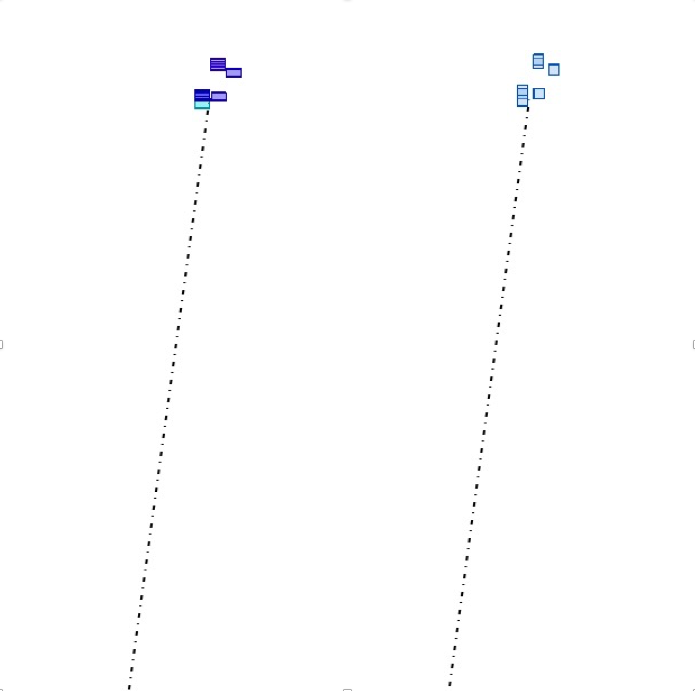}}
\hfill
  \subfigure[1 GeV photon]{\includegraphics[width=0.495\textwidth]{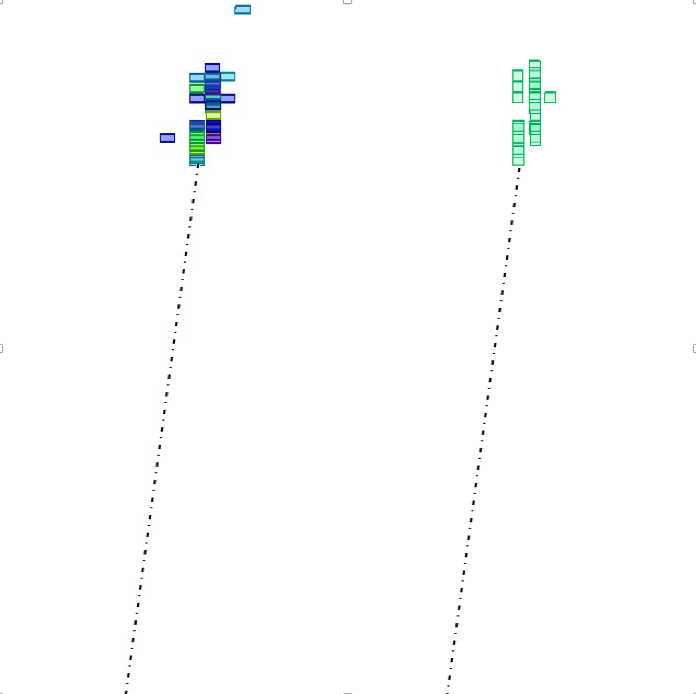}}
\\
  \subfigure[10 GeV photon]{\includegraphics[width=0.495\textwidth]{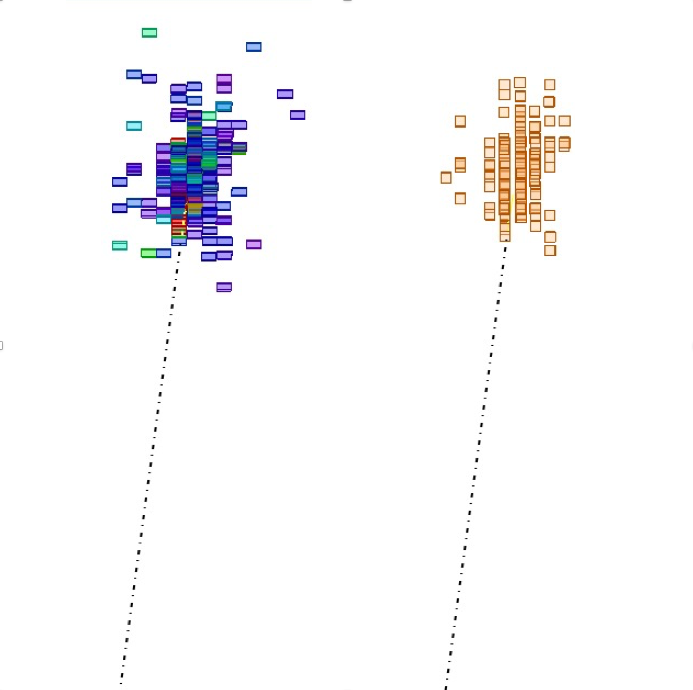}}
\hfill
  \subfigure[100 GeV photon]{\includegraphics[width=0.495\textwidth]{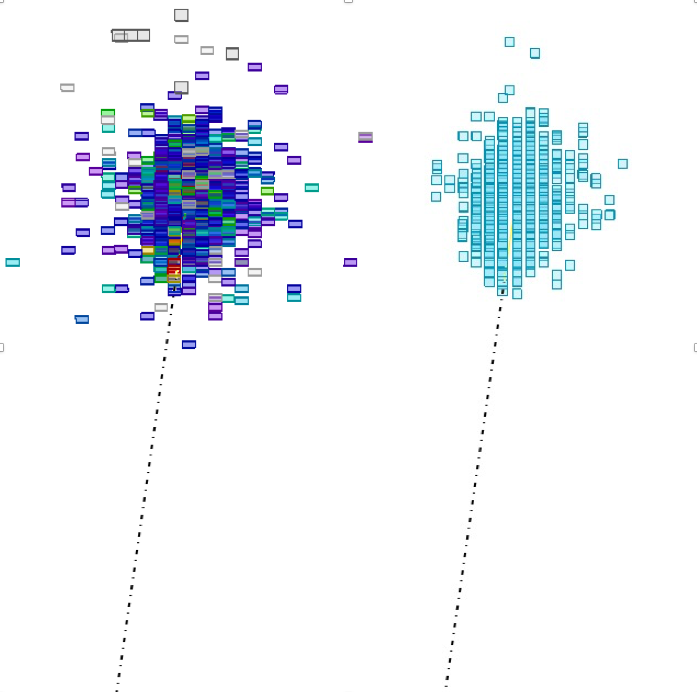}}

\end{minipage}
\caption{ The calorimeter hits and the corresponding reconstructed calorimeter clusters of 0.1 GeV, 1 GeV, 10 GeV and 100 GeV photon.}
\label{fig:photopo}
\end{figure}

Figure~\ref{fig:photopo} shows some calorimeter hits and the corresponding reconstructed calorimeter clusters of 0.1 GeV, 1 GeV, 10 GeV and 100 GeV photons at CEPC baseline detector. The cluster reconstruction efficiency reaches 100\% for photons with energy larger than 200 MeV and decreases to 85\% when the photon energy is 100 MeV. Every neutral cluster passing loose shower shaper requirements is identified as a photon candidate. The photon identification and reconstruction efficiency are discussed in detail in Section~\ref{sec:sinpho}.

%=======samples===========
\subsection{Samples} 
To study the adequate single-photon performance, we simulate single-photon samples over an energy range of 0.1-175 GeV (0.1, 0.2, 0.3, 0.4, 0.5, 0.6, 0.7, 0.8, 0.9, 1, 2, 3, 4, 5, 6, 7, 8, 9, 10, 20, 30, 40, 50, 75, 100, 175 GeV). The single neutron samples over a small energy range of 1-175 GeV with the same binning are also simulated for the photon identification. At each energy point, 10k events are simulated and reconstructed. All the samples follow a flat distribution in $\theta$ and $\phi$ over the 4$\pi$ solid angle. The $\nu\nu{H}$ ($H\to\gamma\gamma$) are simulated to benchmark the photon reconstruction performance.

In addition, single $\pi^{0}$ samples with a flat distribution in energy from 0-50 GeV are simulated and reconstructed to study the di-photon reconstruction efficiency.  The statistic is about 250k. The inclusive Higgs and $Z\to\tau\tau$ samples are also simulated to provide a benchmark for photon reconstruction and separation ability for di-particles.

%++++++++++++++++++++++Results of single photons+++++++++++++++++++++++++++++++

\section{Performance on single-photon events}
\label{sec:sinpho}

\subsection{Photon conversion rate}
\begin{figure*}
\begin{minipage}{\textwidth}
\centering
 \subfigure[]{\includegraphics[width=0.48\textwidth]{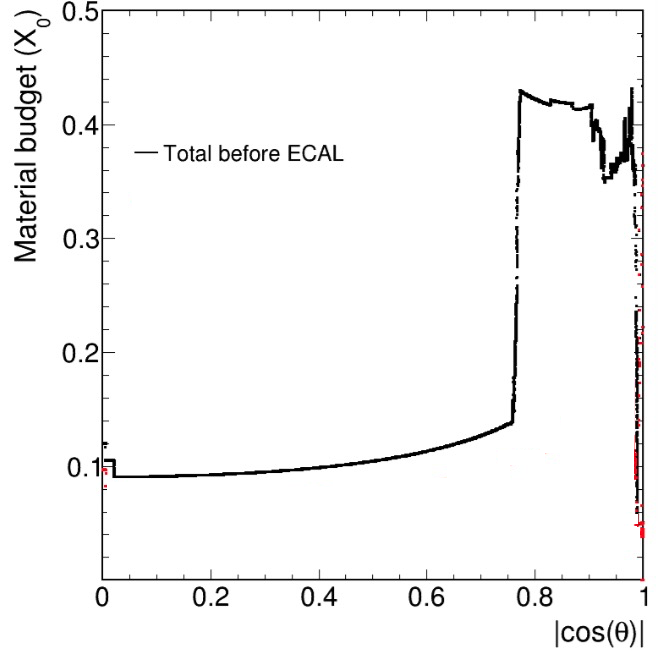}}
\hfill
  \subfigure[]{\includegraphics[width=0.496\textwidth]{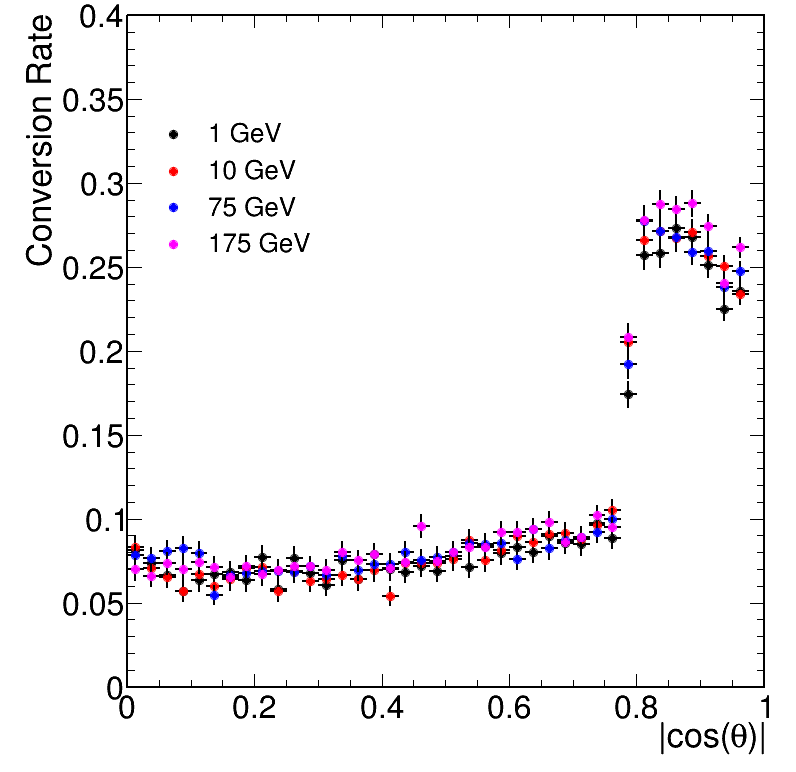}}
\end{minipage}
\caption{(a) shows the material budget in front of ECAL. The material budget in the forward origin is higher than that in the central region, because of the TPC endcap. (b)  shows the photon conversion rate with different energies. Roughly 6-10\% of photons in the central region and 25\% of photons in the forward region will convert to $e^{+}e^{-}$ pairs.}
\label{fig:convert}
\end{figure*}

Photons may convert to $e^{+}e^{-}$ pairs when they interact with the materials in front of ECAL.  Figure~\ref{fig:convert} (a) shows the material budget before ECAL at the CEPC baseline detector. The material budget in the forward origin is higher than that in the central region, because of the TPC endcap. Figure~\ref{fig:convert} (b) shows the photon conversion rate with different energies. The photon conversion rate is observed to be proportional to the material budget and does not depend on the energy significantly. Roughly 6-10\% of photons in the central region and 25\% of photons in the forward region will convert to $e^{+}e^{-}$ pairs~\cite{cdr}. 

%++++++++++++++>>

\subsection{Photon identification and reconstruction efficiency}

Photons and neutral hadrons are reconstructed as neutral clusters by Arbor. Identifying the photons from neutral clusters is crucial for the successful photon reconstruction. Of cause, the merged $\pi^0$ is one major source of the photon misidentification in high energy. The detector design and optimization aim at sufficient control of this effect. The $\pi^0$ reconstruction is discussed in Section~\ref{sec:dipho}. In this section,  the discrimination of photons and neutrons is studied. In addition, the full combined clustering reconstruction and identification efficiency, or photon reconstruction efficiency for short, is also discussed. 

The photon clusters can be identified from the neutral clusters using several discriminating variables~\cite{id1}\cite{id2}. These variables characterize the lateral and longitudinal shower development and the shower leakage fraction in the HCAL, such as the fractal dimension of a shower, the depth of the cluster position, the ratio of the energy deposited in ECAL and HCAL and so on.
%Such variables  The high granularity calorimeter of the CEPC baseline preserves detailed information on the cluster spatial configuration, which is critical for the photon identification. The spatial variables include the the fractal dimension of the and the sine of cluster size. .......adding defined. dan's paper. after meeting
%The photon clusters can be identified from all the neutral clusters using its cluster shape information.  
%CEPC physics reach requires good reconstruction of low energy  photon. The low energy neutron deposits in the ECAL. 

Besides, the CEPC detector assumes a Time-of-Flight (ToF) with 50 ps time resolution~\cite{tof1}\cite{tof2}]\cite{tof3}. This ToF is crucial for the identification of low energy photons. CEPC physics reach requires a good reconstruction of low energy photons. Since neutral hadrons travel slower than photons, the $\Delta{T}$ defined as the difference between the neutral hadron flight time and the photon flight time can be used to separate photons from neutral hadrons. In physics, the $\Delta{T} $ can be calculated as follows:

%For instance, the measurement of $Br(\tau\to{X})$ from $Z\to\tau\tau$ events.
%Using both cluster shape and ToF information, the CEPC reconstruction reaches a high efficiency, high purity photon identification at the baseline geometry. 
$$
\Delta{T} =\frac{L}{\beta{c}}-\frac{L}{c}    
	       = \frac{L}{c}((1-(\frac{m}{E})^2)^{-\frac{1}{2}}-1)
$$
$L$ is the distance the particle travels. Considering the detector geometry, $L$ should be larger than 1.8 m and smaller than 3 m.  For simplicity, we use the average distance, 2.4 m, for later discussion. $c$ is $3\times10^8$ m/s. $m$ is the mass of neutral hadrons. For neutrons, $m$ is 940 MeV. $E$ is true energy. 
Setting a benchmark cluster time resolution of 50 ps, photons can be efficiently separated from neutrons for energy smaller than 8 GeV.
%In physics,

%++++++++++++++>>
Table~\ref{symbols} is the photon identification efficiency. For unconverted photons with energies larger than 1 GeV, the photon identification efficiency is higher than 99\% and the misidentification rate is smaller than 1\%.

\begin{table}
\centering
\caption{Photon identification efficiency in different energy categories}
\begin{tabular}{ccc}
  \hline
 True Energy &  Photon &  Neutron \\

   E$\le$2 GeV &  99.62\%$\pm$0.05\% & 0.00\%$\pm$0.00\% \\

2 GeV$<E\le$5 GeV &   99.92\%$\pm$0.03\% &  0.02\%$\pm$0.00\% \\

5 GeV$<E\le$10 GeV  & 99.93\%$\pm$0.01\% & 0.04\%$\pm$0.00\% \\

$E>10$ GeV  &   99.81\%$\pm$0.02\%  & 0.83\%$\pm$0.05\%\\

  \hline
\end{tabular}
\label{symbols}
\end{table}

\begin{figure*}
\begin{minipage}{\textwidth}
\centering
\subfigure[]{\includegraphics[width=0.495\textwidth]{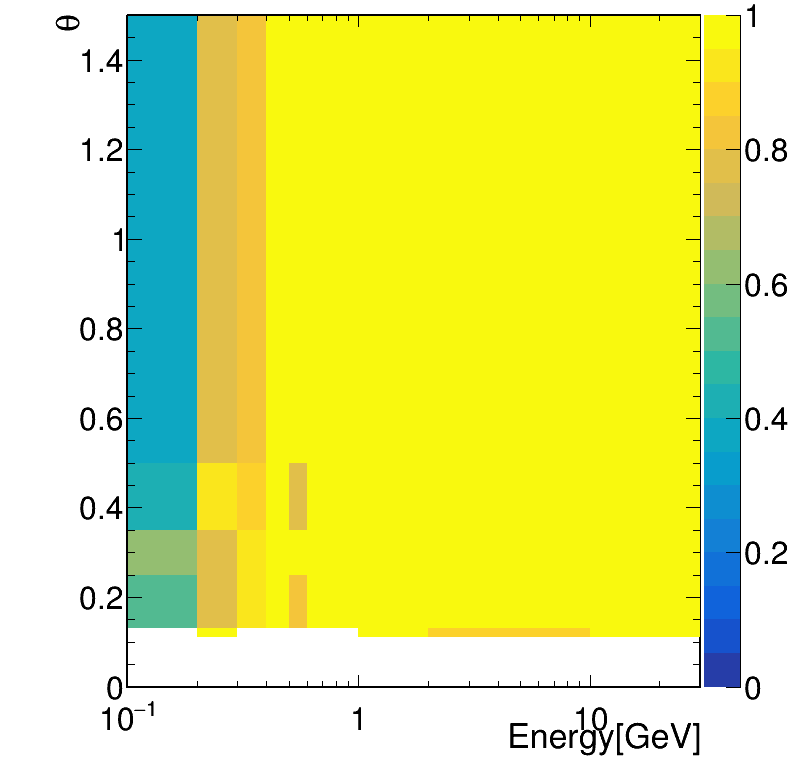}}
\hfill
\subfigure[]{\includegraphics[width=0.495\textwidth]{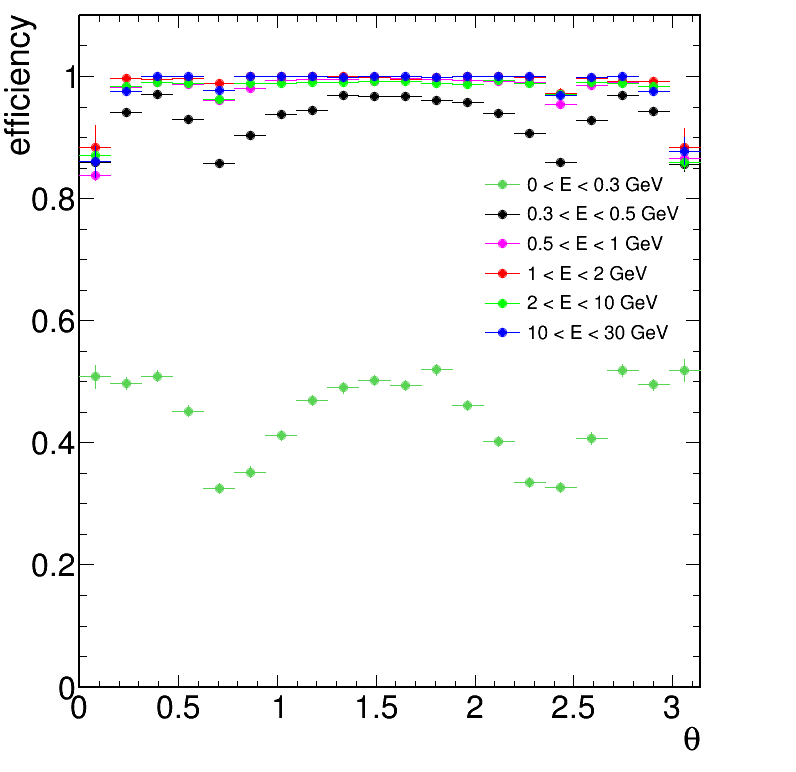}}
\end{minipage}
\caption{(a) is the ($\theta$, Energy) dependences of photon reconstruction efficiency. Between 200 MeV and 500 MeV, efficiencies are varying from 70\% to 99\%. The efficiencies are reaching 99\% when $E>500$ MeV. (b) is the photon reconstruction efficiency as a function of $\theta$ with different energies. The photon reconstruction efficiency is sensitive to the dead zone between the ECAL barrel and endcaps. }
\label{fig:receff}
\end{figure*}

The photon reconstruction efficiency is defined as the probability of the identified photon candidates in all simulated unconverted events. Figure~\ref{fig:receff} (a) shows the ($\theta$, Energy) dependences of photon reconstruction efficiency. Between 200 MeV and 500 GeV, efficiencies are varying from 70\% to 99\% and they are reaching 99\% when $E>500$ MeV. Figure~\ref{fig:receff} (b) is the photon reconstruction efficiency as a function of $\theta$ with different energies. The photon reconstruction efficiency is sensitive to the dead zone between the ECAL barrel and endcaps. 

 As for the reconstruction of converted photons, thanks to the lepton identification performance and the large solid angle coverage,  about 80\% of the converted photon can be reconstructed using a simplistic algorithm~\cite{cdr}. In the future, a dedicated converting photon finding algorithm will be developed.

% \begin{table}[!h]

\subsection{Photon energy resolution}

\begin{figure}
\begin{minipage}{\textwidth}
\centering
\includegraphics[width=0.495\textwidth]{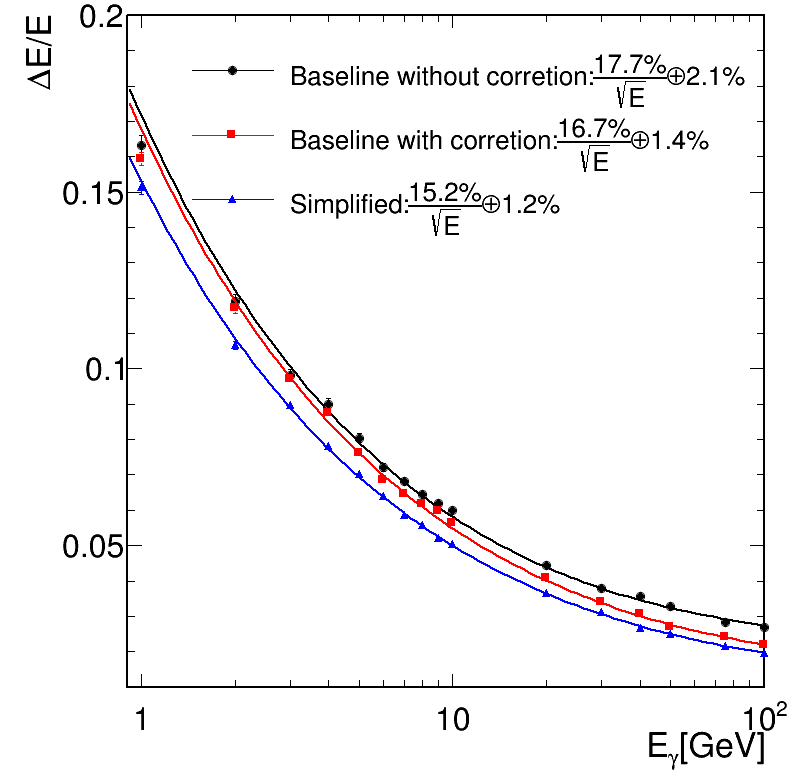}
\end{minipage}
\caption{The photon energy resolution as a function of energy in the central region.}
\label{fig:corr1}
\end{figure}

\begin{figure*}
\begin{minipage}{\textwidth}
\centering
\subfigure[]{\includegraphics[width=0.495\textwidth]{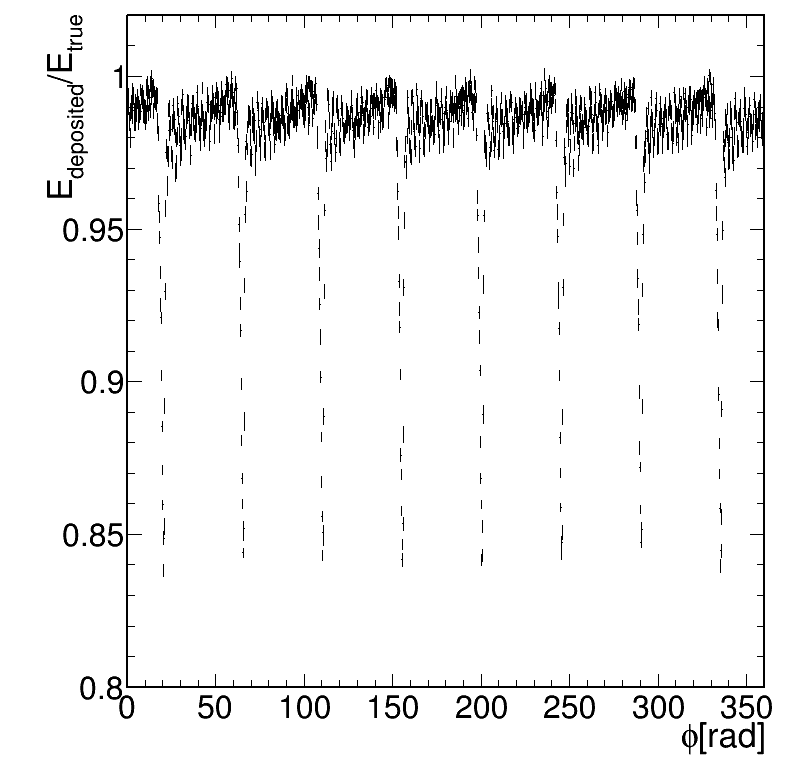}}
\hfill
\subfigure[]{\includegraphics[width=0.495\textwidth]{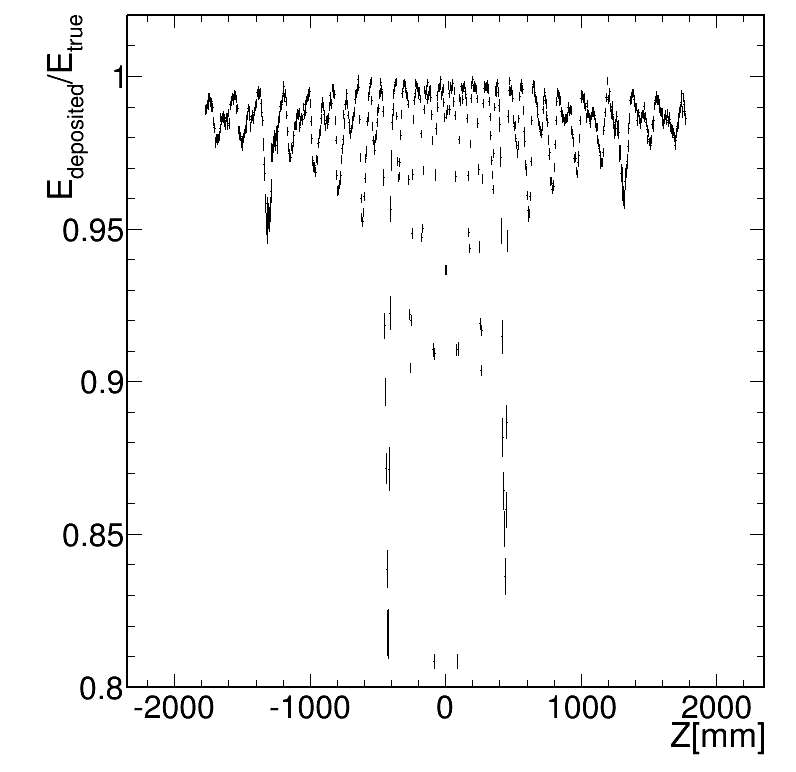}}
\end{minipage}
\caption{The $\frac{E_{deposited}}{E_{true}}$ distributions as a function of $\Phi$ (a) and $Z$ (b) with 50 GeV photons in the central region. They reflect the detailed geometry structure of the baseline detector.} 
\label{fig:corr2}
\end{figure*}

\begin{figure*}
\begin{minipage}{\textwidth}
\centering
\subfigure[]{\includegraphics[width=0.495\textwidth]{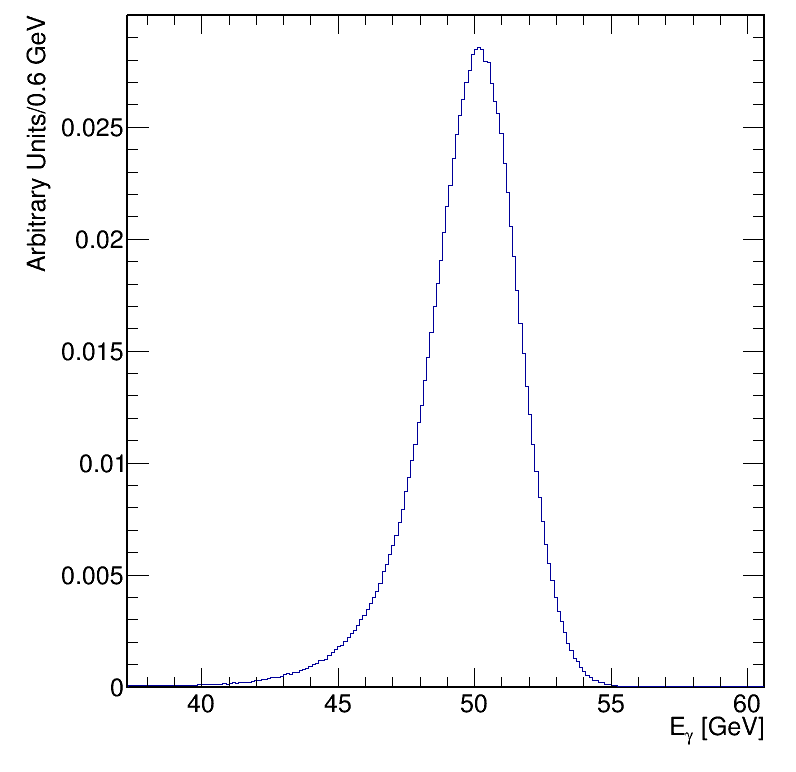}}
\hfill
\subfigure[]{\includegraphics[width=0.495\textwidth]{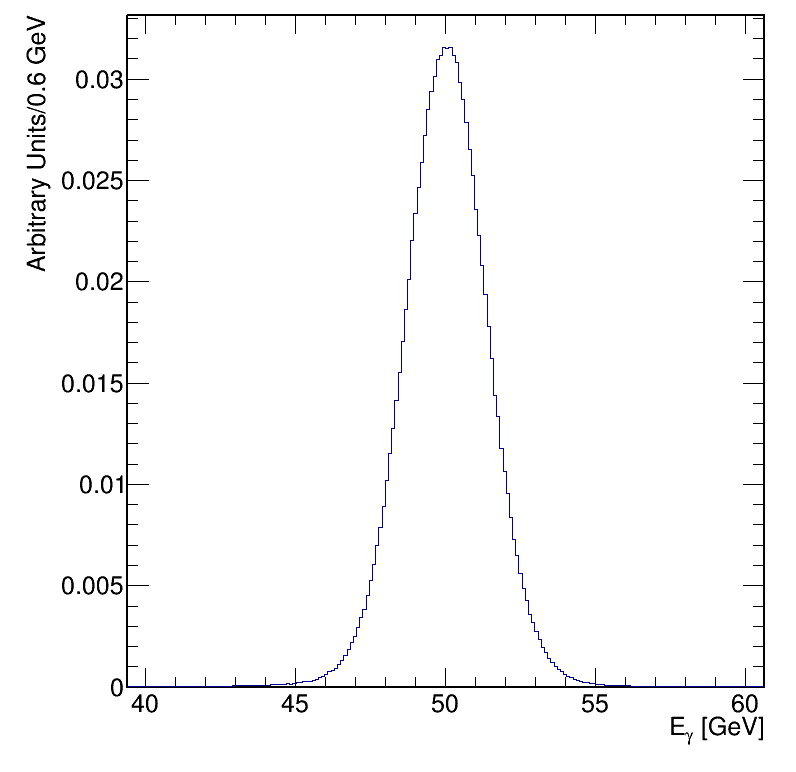}}
\end{minipage}
\caption{The energy distributions of 50 GeV photon before (a) and after (b) applying the energy correction algorithm. Roughly 1.5 million events, normalized to unit area.}
\label{fig:50gev}
\end{figure*}

The photon energy resolution is sensitive to the tracker material and calorimeter geometry defects, such as the cracks between the ECAL modules, staves, and the dead zone between the ECAL barrel and endcaps. To quantify their impact, a simplified geometry~\cite{zhaopfa} is implemented. The simplified geometry is organized into a cylindrical barrel and two endcaps, forming a closed cylinder. The simplified geometry has no geometry defects and no materials before ECAL and uses the same ECAL geometry parameters as the CEPC baseline detector. In this study, only unconverted photon events are used. Figure~\ref{fig:corr1} compares the energy resolution of unconverted photons at the baseline detector with the resolution at the simplified geometry in the central region ($|cos\theta|<0.7$). The resolution of simplified geometry represents the ultimate resolution of the detector and is consistent with the CALICE prototype test beam result~\cite{calice}. The degradation in the resolution of the baseline detector is caused by the material in the tracker and geometric inhomogeneities.

The geometry-based correction algorithm has been developed to scales the EM clusters located at the geometry cracks. The correction of the geometry defects at any realistic detector geometry is vital for the photon reconstruction. The corrected energy is estimated as follows.
 $$
 E_{deposited}^{corrected}=\frac{E^{'}_{true}}{E^{'}_{deposited}}\times{E_{deposited}}
 $$
In the equation, the scale factor $\frac{E^{'}_{true}}{E^{'}_{deposited}}$ is obtained from the simulated 50 GeV photon samples.  Figure~\ref{fig:corr2} shows the $\frac{E_{deposited}}{E_{true}}$ distribution as a function of $\Phi$ and $Z$ in the central region. It reflects the detailed geometry structure of the baseline detector which is discussed in Section~\ref{sec:detectorsoft}. %Since the junction of two staves is asymmetry, there is a small increase in the energy response with $\phi$ in each stave. In addition, the bigger boundary effects at Z$\approx$400 mm is because the gap between two modules is bigger than the gap between two columns.%the photons in that region are more likely to hit into the gap directly.

Figure~\ref{fig:50gev} shows the energy distribution of 50 GeV photon before and after applying the energy correction algorithm.
Without energy correction, the energy distribution has a long tail caused by the geometry defects. Then the effective energy resolution is  3.4\%.  After applying the energy correction, the energy resolution is 2.7\%. 

\begin{figure}
\begin{minipage}{\textwidth}
\centering
\includegraphics[width=0.495\textwidth]{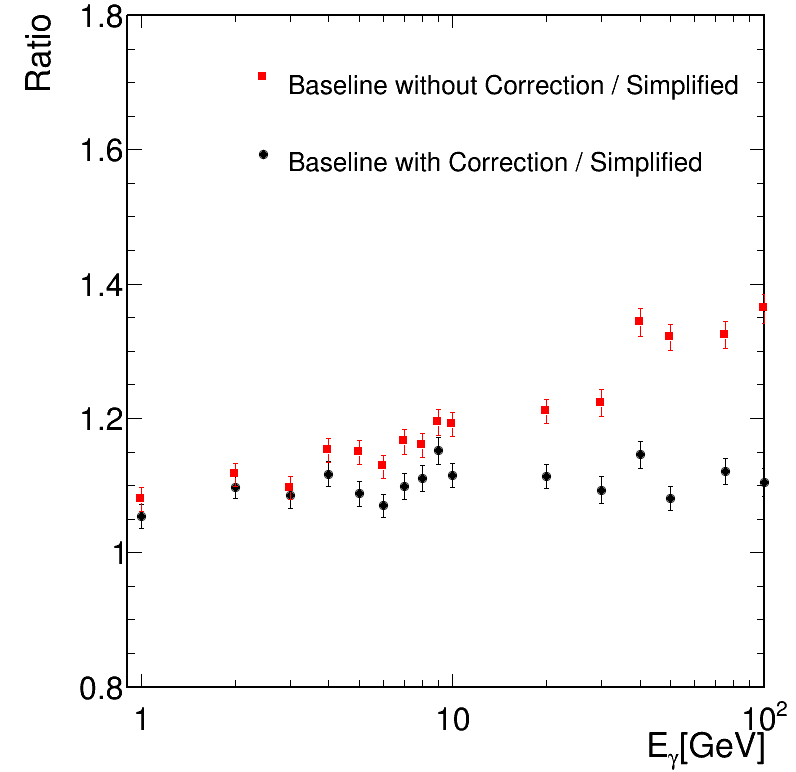}
\end{minipage}
\caption{The ratio of the energy resolution at CEPC baseline detector to the resolution at simplified geometry. Because the input sample is chosen at 50 GeV, correction at high energy is more significant.}
\label{fig:ratio}
\end{figure}

The energy resolution as a function of energy with correction is showed in Figure~\ref{fig:corr1} in the red line. Figure~\ref{fig:ratio} shows the ratio of the energy resolution at the CEPC baseline detector to the resolution at simplified geometry. Because the input sample is chosen at 50 GeV, correction at high energy is more significant. The energy-dependent correction algorithm will be developed for later analysis.

%The photon energy resolution benchmarked by the diphoton mass distribution will be discussed in next section.

%++++++++++++++++++++++Results of Di photons+++++++++++++++++++++++++++++++
\section{Performance on di-photon events}
\label{sec:dipho}

\subsection{$H\to\gamma\gamma$}
Successful photon reconstruction is essential for precise measurements of the Higgs boson. In this study, we benchmark the photon reconstruction using the Higgs mass resolution with $\nu\nu{H}$ ($H\to\gamma\gamma$) sample.

\begin{figure}
\begin{minipage}{\textwidth}
\centering
\subfigure[]{\includegraphics[width=0.495\textwidth]{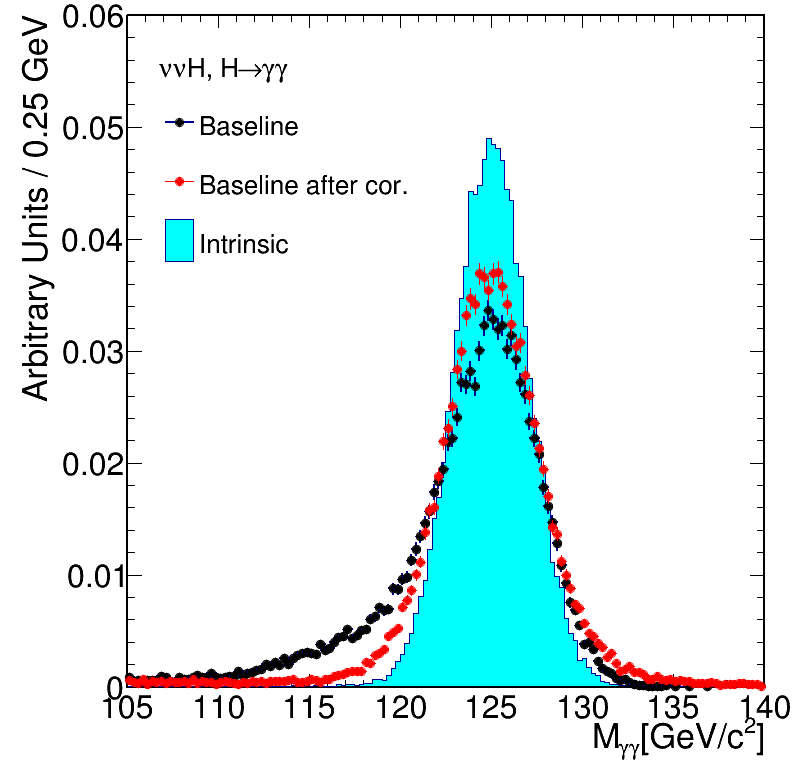}}
\end{minipage}
\caption{The Higgs boson invariant mass reconstructed from $H\to\gamma\gamma$ events in the central region. A relative mass resolution of 1.7\% is achieved at the simplified geometry. At baseline detector, the mass resolution is about 2.6\%. After applying the energy correction algorithm, the low energy tail is eliminated and the mass resolution is approximately 2.2\%. Roughly 6k events, normalized to unit area.} 
\label{fig:higgsmass}
\end{figure}

Figure~\ref{fig:higgsmass} shows the Higgs boson invariant mass reconstructed in the central region. At the simplified geometry, the relative mass resolution is approximately 1.7\%. At the baseline detector, the Higgs boson invariant mass distribution has a low energy tail caused by the geometry defects. The mass resolution is about 2.6\%. After applying the energy correction algorithm discussed in the last section, the low energy tail is eliminated.  A relative mass resolution of 2.2\% is achieved, which is consist with the CALICE prototype test beam result~\cite{calice}. The resolution is expected to improve significantly once the energy-dependent correction algorithm are implemented.

\begin{figure*}
\begin{minipage}{\textwidth}
\centering
\subfigure[]{\includegraphics[width=0.495\textwidth]{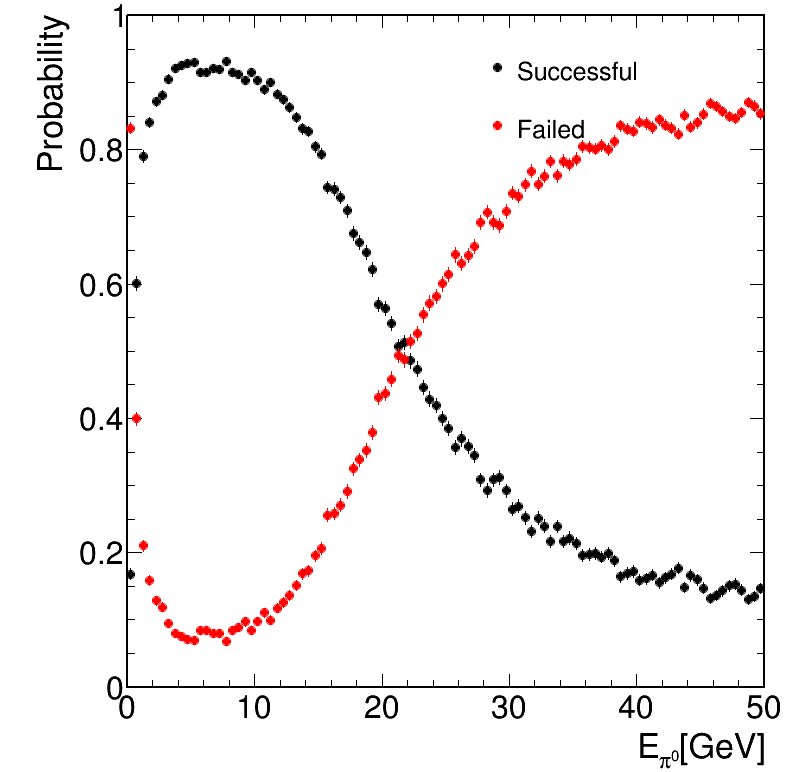}}
\hfill
\subfigure[]{\includegraphics[width=0.495\textwidth]{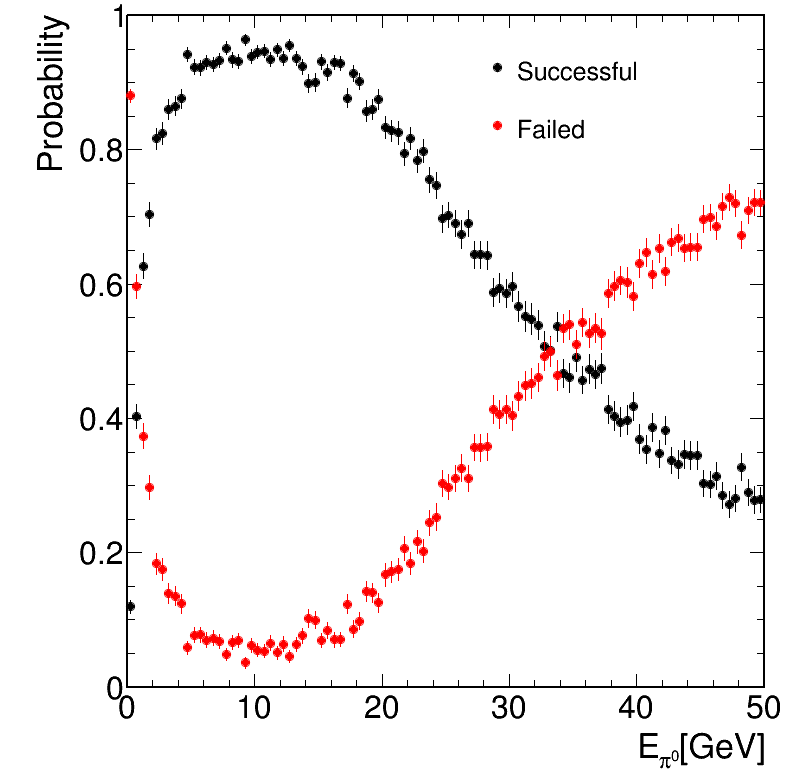}}
\end{minipage}
\caption{The probability of successfully reconstructing $\pi^{0}$ in the barrel region (a, $|cos\theta|<0.8$) and in the endcap region (b, $|cos\theta|>0.8$) at defect-free ECAL  geometry. In the barrel region, above 50\% of $\pi^0$ can be reconstructed when the energy of $\pi^0$ is lower than 22 GeV. While in the endcap region, a better than 50\% reconstruction rate is observed once the energy of $\pi^0$ is lower than 34 GeV. The lower reconstruction efficiency at the low energy is caused by the effects of photon identification and reconstruction.}
\label{fig:pionrecoeff}
\end{figure*}

\begin{figure}
\begin{minipage}{\textwidth}
\centering
\subfigure{\includegraphics[width=0.495\textwidth]{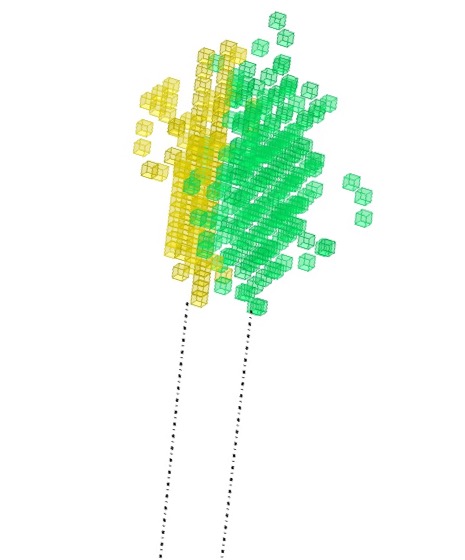}}
\end{minipage}
\caption{A successfully reconstructed 19 GeV $\pi^0$. The calorimeter showers are close to each other but can be separated. } 
\label{fig:pion0}
\end{figure}

\begin{figure*}
\begin{minipage}{\textwidth}
\centering
\subfigure[]{\includegraphics[width=0.495\textwidth]{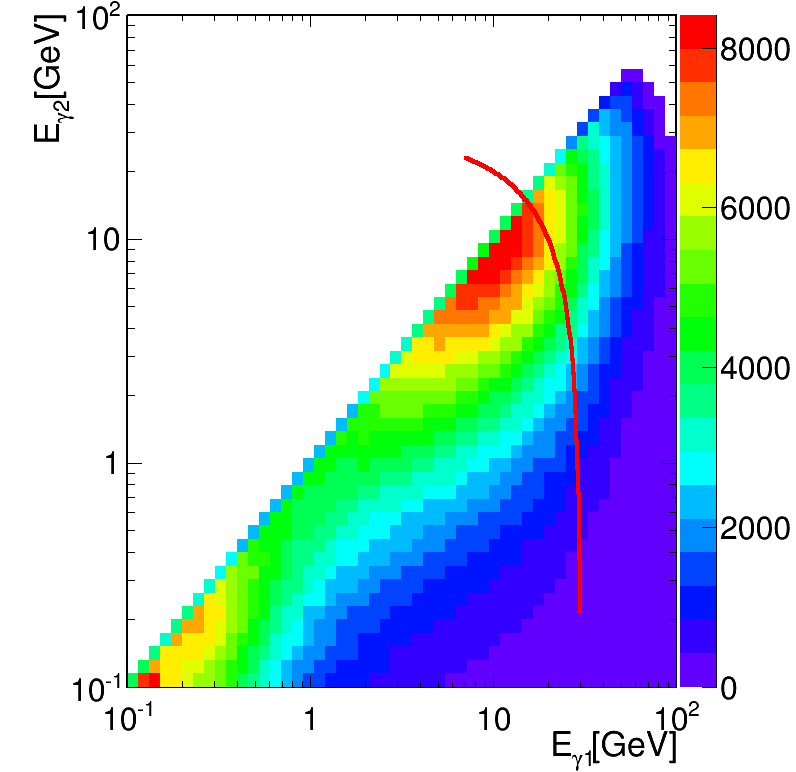}}
\hfill
\subfigure[]{\includegraphics[width=0.495\textwidth]{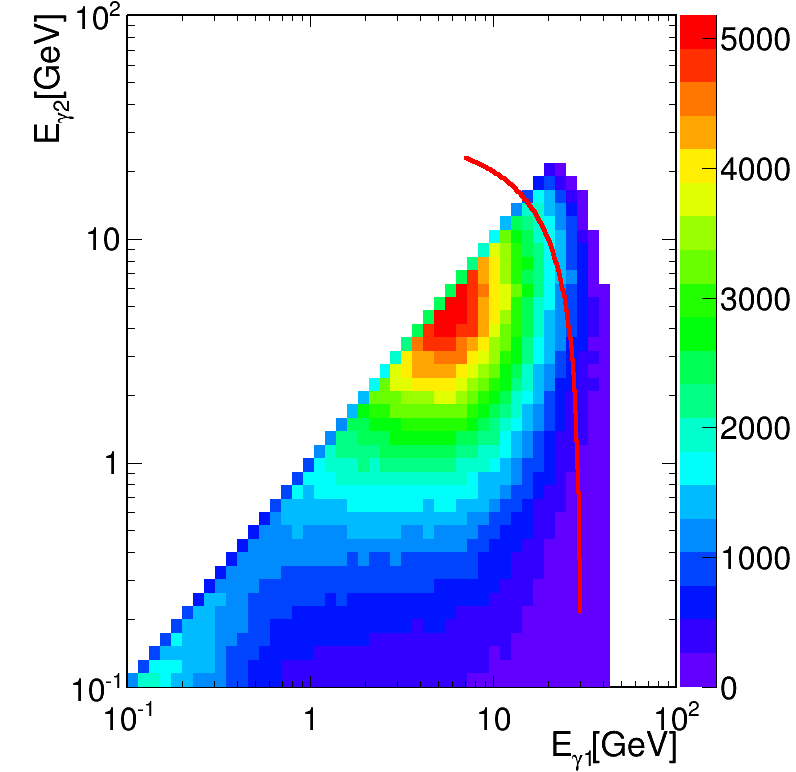}}
\end{minipage}
\caption{The generated $\pi^{0}$ distribution as a function of the energies of di-photons from  $\pi^{0}\to\gamma\gamma$ in inclusive Higgs (a) and $Z\to\tau\tau$ samples (b). $E_{\gamma1}$ is the energy of the leading photon. $E_{\gamma2}$ is the energy of the sub-leading photon. The red line is the function of $E_{\gamma1} + E_{\gamma2}$ = 30 GeV.}
\label{fig:pion}
\end{figure*}

\subsection{$\pi^{0}\to\gamma\gamma$}
The separation performance of nearby photon clusters can be characterized by $\pi^0$ reconstruction. The $\pi^0$ itself is an important physics object for $\tau$ identification. In this section, the $\pi^{0}$ reconstruction performance is discussed.

%The di-photon separation ability depends mainly on the ECAL cell size and the separation distance. For $10\times 10$ mm$^2$ cell size, the critical distance is 16 mm~\cite{s5}\cite{zhaopfa}. 
We simulate single $\pi^{0}$ events to estimate the probability of successfully reconstructing $\pi^{0}$ as a function of energy. The probability of successfully reconstructing $\pi^{0}$ is defined as the probability of successfully reconstructing two photons at least and with the leading invariant mass between (0.135-5$\sigma$, 0.135+5$\sigma$) MeV. The $\sigma$ is calculated as follows:
 $$\frac{\sigma}{m} = \frac{1}{2}(\frac{\sigma_{E_1}}{E_1}\oplus\frac{\sigma_{E_2}}{E_2})$$
In this equation, $\frac{\sigma_{E_1}}{E_1}$ and $\frac{\sigma_{E_1}}{E_2}$ are the energy resolutions of the leading photon and the sub-leading photon.
$m$ is the invariant mass of $\pi^{0}$.

The Figure~\ref{fig:pionrecoeff} shows the probability of successfully reconstructing $\pi^{0}$ at the barrel and endcap region. In the barrel region, above 50\% of $\pi^0$ can be reconstructed when the energy of $\pi^0$ is lower than 22 GeV . While in the endcap region, a better than 50\% reconstruction rate is observed once the energy of $\pi^0$ is lower than 34 GeV. In general, the reconstruction efficiency in the endcap is higher than the efficiency in the barrel region. The lower reconstruction efficiency at the low energy is caused by the effects of photon identification and reconstruction. Considering the geometry parameters, those critical energies are consistent with the ECAL separation performance study in Ref~\cite{zhaopfa}\cite{zhaopfa2}. For simplicity, we use the 30 GeV as the critical energy for later discussion. 

The Figure~\ref{fig:pion0} is a plot of a successfully reconstructed $\pi^0$. The calorimeter showers are close to each other but can be separated.

%=====pi0 in  higgs Z samples======
The Figure~\ref{fig:pion} shows the distribution of $\pi^{0}$ events generated in the inclusive Higgs (a) and $Z\to\tau\tau$ samples (b) at CEPC baseline detector. $E_{\gamma1}$ is the energy of the leading photon.  $E_{\gamma2}$ is the energy of the sub-leading photon. The red line represents the critical energy of 30 GeV.  In the inclusive Higgs sample, about 15\% of $\pi^0$  has energy above 30 GeV. Only  3\% of  $\pi^0$ in $Z\to\tau\tau$ sample has its energy larger than the critical energy. So most of the $\pi^{0}$ can be reconstructed successfully.

%++++++++++++++++++++++conclustion++++++++++++++++++++++++++++++

\section{Conclusion}
\label{sec:con}
The photon reconstruction is a critical physics performance for the CEPC. Using the baseline CEPC simulation tool, we analyze the photon reconstruction performance at the CEPC baseline detector. The photon conversion rate, the reconstruction/identification performance, and the energy resolution are analyzed at the single-photon samples. We also quantify the Higgs mass resolution at $H\to\gamma\gamma$ event, and the $\pi^0$ reconstruction performance with corresponding full simulation samples. 

At single-photon samples, a high efficient reconstruction/identification performance is observed, while the conversion rate is observed to be consistent with the tracker materials. Using the cluster shape and ToF information, the photon identification efficiency is higher than 99\% and the misidentification rate is smaller than 1\% for the unconverted, isolated photons with energies larger than 1 GeV.

The photon energy resolution is analyzed at both the CEPC baseline detector and a simplified, defect-free ECAL geometry. The latter is used as a reference to study the impact of geometry defects (i.e., gaps and dead zones between the ECAL modules) on the photon energy resolution. The photon energy can be measured to an accuracy of  $\frac{17.7\%}{\sqrt{E}}\oplus2.1\%$/$\frac{15.2\%}{\sqrt{E}}\oplus1.2\%$ at the baseline/simplified geometry. At the $H\to\gamma\gamma$ samples, the relative mass resolution of the Higgs boson is 2.2\%/1.7\%. 

We observed a significant impact of the geometry defects on the low energy tail and photon energy resolution. The relative degradation (baseline/simplified) is observed to be proportional to the photon energy and can be as large as 35\% at photons with energy between 40 and 100 GeV. On the other hand, this degradation can be ameliorated by a photon-position based correction algorithm. As a test of principle, we developed a prototyping correction algorithm that uses a 50 GeV photon sample as a reference, this algorithm could efficiently eliminate the low energy tail induced by the geometry defects, and reduce the relative degradation from 35\% to 10\%. Consistent improvement is observed also in $H\to\gamma\gamma$. Since the correction algorithm is sensitive to the photon energy, an iterative correction algorithm shall be developed in the future. 

The $\pi^0$ reconstruction performance is limited by the photon energy threshold of ECAL, the geometry acceptance, and the merging of photon clusters at high energy. 
We analyze the $\pi^0$ reconstruction efficiency as a function of the $\pi^0$ energy, the critical energy of 22/34 GeV at the barrel/endcap region (corresponding to 50\% of the successful reconstruction rate) is observed using baseline reconstruction (Arbor). Giving the geometry parameters, that critical energy is consistent with the ECAL separation performance study from Ref~\cite{zhaopfa}\cite{zhaopfa2}.

For simplicity, the average critical energy of 30 GeV is used for later discussion.
At 240 GeV center of mass energy, roughly 15\% of the $\pi^0$ generated in the inclusive Higgs sample has its energy above the average critical energy. More importantly, only 3\% of the  $\pi^0$ generated in $Z\to\tau\tau$ events exceeds this critical energy threshold. It should be remarked that Arbor is a general PFA reconstruction, and this $\pi^0$ reconstruction efficiency and the separation performance could be enhanced by applying a dedicated $\pi^0$ finding and identification algorithm. 

To conclude, the CEPC baseline detector reconstructs the isolated photons with high purity and high efficiency. It provides efficient separation power, such that most of the $\pi^0$ generated in objective physics events can be successfully reconstructed. In terms of energy resolution, it achieves a relative mass resolution of 2.2\%, which is foreseen to be improved significantly with a dedicated position-dependent correction algorithm.
This Particle Flow oriented detector fulfills the physics requirements on the photon reconstruction for the CEPC Higgs/EW measurements and serves as a reasonable starting point for future optimization.

\section*{Acknowledgement}
%\begin{acknowledgements}
We are grateful to Hang Zhao, Dan Yu, Zhigang Wu and Hao Liang for their supports and helps.
We thank Gang Li and Chengdong Fu for the physics event generator files. 

This work was supported by National Key Program for
S\&T Research and Development (Grant No. 2016YFA0400400), the
National Natural Science Foundation of China (Grant Nos. 11675202, 11705016), the Beijing Municipal Science \& Technology Commission project (Gran No. Z1811000042180043), the Natural Science Foundation of the Jiangsu Higher Education Institutions of China (Grant No. 17KJB140002), the key project of Natural Science Foundation of the Changzhou Institute of Technology (Grant No. YN1629).
%\end{acknowledgements}


\begin{thebibliography}{4}
\bibitem{a}
ATLAS Collaboration, \emph{Observation of a new particle in the search for the Standard Model Higgs boson with the ATLAS detector at the LHC}, Physics Letters B, {\bf 716:} 1-29 (2012), arxiv: 1207.7214 [hep-ex]
\bibitem{b}
CMS Collaboration, S. Chatrchyan et al., \emph{Observation of a new boson at a mass of 125 GeV with the CMS experiment at the LHC}, Physics Letters B, {\bf 716:} 30-61 (2012), arXiv: 1207.7235 [hep-ex]

\bibitem{sm1}
C. D. Froggatt and H. B. Nielsen, \emph{Hierarchy of Quark Masses, Cabibbo Angles and CP Violation}, Nucl. Phys. B, {\bf 147:} 277-298 (1979).
\bibitem{sm2}
P. A. M. Dirac, \emph{New basis for cosmology}, Proc. Roy. Soc. Lond. A165 (1938).
\bibitem{sm3}
G.$^{'}$t Hooft, \emph{Naturalness, chiral symmetry, and spontaneous chiral symmetry breaking}, NATO Sci. Ser. B, {\bf 59:} 135-157 (1980).
\bibitem{sm4}
R. Essig, P. Meade, H. Ramani, and Y.-M. Zhong, \emph{Higgs-Precision Constraints on Colored Naturalness}, JHEP, {\bf 09:} 085 (2017), arXiv:1707.03399 [hep-ph].
\bibitem{sm5}
C. Cai, Z.-H. Yu, and H.-H. Zhang, \emph{CEPC Precision of Electroweak Oblique Parameters and Weakly Interacting Dark Matter: the Fermionic Case}, Nucl. Phys. B, {\bf 921:} 181-210 (2017), arXiv:1611.02186 [hep-ph].
\bibitem{sm6}
Q.-F. Xiang, X.-J. Bi, P.-F. Yin, and Z.-H. Yu, \emph{Exploring Fermionic Dark Matter via Higgs Boson Precision Measurements at the Circular Electron Positron Collider}, Phys. Rev. D, {\bf 97:} 055004 (2018), arXiv:1707.03094 [hep-ph].


\bibitem{cdr}
The CEPC-SPPC Study Group, \emph{CEPC Conceptual Design Report: Volume 2 - Physics \& Detector}, arXiv: 1811.10545 [hep-ex]
\bibitem{zl}
Z. Liu, L.-T. Wang, and H. Zhang, \emph{Exotic decays of the 125 GeV Higgs boson at future $e^+e^-$  lepton colliders}, Chin. Phys. C, {\bf 41} 063102 (2017), arXiv:1612.09284 [hep-ph].


\bibitem{s1}
\emph{CEPC software website}, \url{http://cepcsoft.ihep.ac.cn/}

\bibitem{soft1}
W. Kilian, T. Ohl, and J. Reuter, \emph{WHIZARD: Simulating Multi-Particle Processes at LHC and ILC},  Eur. Phys. J. C, {\bf 71:} 1742 (2011), arXiv:0708.4233 [hep-ph]
%\bibitem{soft2}
%J. Alwall, R. Frederix, S. Frixione, V. Hirschi, F. Maltoni, O. Mattelaer, H. S. Shao, T. Stelzer, P. Torrielli, and M. Zaro,  \emph{The automated computation of tree-level and next-to-leading order differential cross sections, and their matching to parton shower simulations}, JHEP, {\bf07:} 079 (2014), arXiv: 1405.0301 [hep-ph]
\bibitem{soft3}
The Pythia Group, \emph{An Introduction to PYTHIA 8.2}, Comput. Phys. Commun., {\bf 191:} 159-177 (2015), arXiv:1410.3012 [hep-ph]
\bibitem{s2}
P. Mora de Freitas and H. Videau, \emph{Detector simulation with MOKKA/GEANT4: Present and future}, LC-TOOL-2003-010
\bibitem{s3}
Source code of MokkaPlus, \url{http://cepcgit.ihep.ac.cn/cepcsoft/MokkaC}
\bibitem{sadd}
\emph{GEANT4 software website}, \url{http://geant4.web.cern.ch/}

\bibitem{s5}
M. Ruan and H. Videau, \emph{Arbor, a new approach of the Particle Flow Algorithm}, arxiv: 1403.4784[physics.ins-det]
\bibitem{s4}
M. Thomson, \emph{Particle Flow Calorimetry and the PandoraPFA Algorithm}, Nucl.Instrum. Meth. A,  {\bf611:} 25-40 (2009), arxiv: 0907.3577  [physics.ins-det]


\bibitem{id1}
CMS collaboration, \emph{Performance of photon reconstruction and identification with the CMS detector in proton-proton collisions at $\sqrt{s}=8TeV$}, JINT, {\bf10:} P08010 (2015), arXiv: 1502.02702
\bibitem{id2}
ATLAS collaboration, \emph{Measurement of the photon identification efficiencies with the ATLAS detector using LHC Run-1 data}, Eur. Phys. J. C, {\bf 76} 666 (2016).








 
\bibitem{tof1}
 CMS Collaboration, \emph{The CMS HGCAL detector for HL-LHC upgrade}, arxiv: 1708.08234[physics.ins-det]
\bibitem{tof2}
 M. Lucia, \emph{A High-granularity Timing Detector for the  Phase-II upgrade of the ATLAS Detector system}, ATL-LARG-SLIDE-2017-008
 \bibitem{tof3}
 F. An, S. Prell, C. Chen, J. Cochran, X. Lou, M. Ruan, \emph{Monte Carlo study of particle identification at the CEPC using TPC $dE/dx$ information}, Eur. Phys. J. C, {\bf 78:} 464 (2018).




\bibitem{calice}
 F. Sefkow et al., \emph{Experimental Tests of Particle Flow Calorimetry.} DESY 14-032, KYUSHU-RCAPP 2015-01, LAL 15-235 (2015), ArXiv: 1507.05893 [physics.ins-det]
 
 
 
\bibitem{zhaopfa}
H. Zhao, C. Fu, D.Yu, Z. Wang, T. Hu and M.Ran, \emph{Particle flow oriented electromagnetic calorimeter optimization for the circular electron positron collider}, JINST, {\bf13:} P03010 (2018), arxiv: 1712.09625v3 [physics.ins-det]
\bibitem{zhaopfa2}
H. Zhao, \emph{Di-photon separation Study and the Higgs Signal at CEPC}, presentation at the Workshop on CEPC 2018





 
 
\end{thebibliography}
\end{document}